\documentclass[10pt]{article}
\usepackage[T1]{fontenc}
\usepackage[english]{babel}
\usepackage[latin2]{inputenc}
\usepackage[margin=2.5cm]{geometry}
\usepackage{graphicx}% Include figure files
\usepackage{subfig}
\usepackage[numbers,compress,sort]{natbib}
\usepackage{soul}
\usepackage{xcolor}
\usepackage{setspace}
 
\title{Modeling of  ultrafast X-ray induced  magnetization dynamics in magnetic multilayer systems}
\author{%
K. J. Kapcia$^{1,2,\dagger}$, V. Tkachenko$^{3,4,\dagger}$,  F. Capotondi$^{5}, $ A. Lichtenstein$^{4,6}$, \\ S. Molodtsov$^{4}$, L. Mueller$^{7}$, A. Philippi-Kobs$^{7}$, P. Piekarz$^{3}$, B. Ziaja$^{1,3,\dagger}$
}
\date{\today}
\begin{document}
\maketitle
{\small{
\begin{flushleft}
$^1$ Center for Free-Electron Laser Science CFEL, Deutsches Elektronen-Synchrotron DESY, Notkestr.  85, 22607 Hamburg, Germany\newline
$^2$ Institute of Spintronics and Quantum Information, Faculty of Physics, Adam Mickiewicz University in Pozna\'n,\\ Uniwersytetu Pozna\'nskiego 2, 61-614 Pozna\'n, Poland\newline 
$^3$ Institute of Nuclear Physics, Polish Academy of Sciences,  Radzikowskiego 152, 31-342 Krak\'ow, Poland\newline
$^4$ European XFEL GmbH, Holzkoppel 4, 22869 Schenefeld, Germany\newline
$^5$ Elettra-Sincrotrone Trieste S.C.p.A, 34149, Trieste, Basovizza, Italy\newline
$^6$ University of Hamburg, Jungiusstr.  9, 20355 Hamburg, Germany\newline
$^7$ Deutsches Elektronen-Synchrotron DESY, Notkestr.  85, 22607 Hamburg, Germany\newline

%$*$ These authors contributed equally to this work.\newline

$\dagger$ Corresponding authors: konrad.kapcia@amu.edu.pl, victor.tkachenko@xfel.eu, ziaja@mail.desy.de 

\end{flushleft}
}}
\begin{abstract} 
\noindent
In this work, we report on modelling results obtained with our recently developed simulation tool enabling nanoscopic description of electronic processes in X-ray irradiated ferromagnetic materials. With this tool, we have studied the response of Co/Pt multilayer system irradiated by an ultrafast extreme ultraviolet pulse at the M-edge of Co (photon energy $\sim$ 60 eV). It was previously investigated experimentally at the FERMI free-electron-laser facility, using the magnetic small-angle X-ray scattering technique. Our simulations show that the magnetic scattering signal from cobalt decreases on femtosecond timescales due to electronic excitation, relaxation and transport processes both in the cobalt and in the platinum layers, following the trend observed in the experimental data. The confirmation of the predominant role of electronic processes for X-ray induced demagnetization in the regime below the structural damage threshold is a step towards quantitative control and manipulation of X-ray induced magnetic processes  on femtosecond timescales. 
\end{abstract}
%%%%%%%%%%%%%%%%%%%%%%%%%%%%%%%%%%%%%%%%%%%%%%%%
\section*{Introduction}
X-ray and extreme ultraviolet (XUV) free-electron lasers (FELs)~\cite{Ackermann,Emma2010,Pile,Allaria,Flavio13} enable investigation of X-ray induced demagnetization within magnetic materials on femtosecond timescales. FELs generate intense, coherent pulses of femtosecond duration and tunable wavelength, which can rapidly induce  strong electronic excitation in solid materials. Historically, since its discovery in 1996 \cite{Beaurepaire}, ultrafast demagnetization on sub-picosecond timescales was studied mostly with lasers working in the infrared wavelength regime~\cite{Kiril10,Koopmans10,Pfau12,Sander17}. X-ray FELs provide not only an opportunity to probe magnetic properties of solids on femtosecond timescales and at nanometer length scales but they also enable to study ultrafast demagnetization induced by photons of much higher energies than those accessible with optical lasers~\cite{Andre21,Schneider2020,Mueller13,Scherz,Wu,Wang2012}. This is possible with  resonant X-ray magnetic scattering  \cite{Hannon98,Hill96,Brueckel}. The energy of photons in the FEL beam is then tuned to an absorption edge of a ferromagnetic element. Transient magnetic properties of the system can be followed, taking advantage of X-ray magnetic circular dichroism (XMCD) effect, for example,  by performing a resonant magnetic small-angle X-ray scattering (mSAXS) measurement \cite{Gutt10,Willems17}. In particular, for samples characterized by a strong perpendicular magnetic anisotropy, the latter scheme gives access to X-ray induced ultrafast changes within  magnetic domains \cite{Riepp,Bran10}. Let us emphasize that in such experiments \cite{Mueller13,Schneider2020,Andre21} the X-ray pulse serves both as a pump and as a probe, exciting the material and simultaneously probing the excited state with magnetic scattering.  

The mSAXS measurement principle is the following (see, e.g., \cite{Gutt10}). The resonant coherent elastic scattering amplitude for a magnetic ion includes a contribution from  charge and magnetic scattering \cite{Hill96,Hannon98}. For an X-ray beam (i) arriving perpendicularly to the surface of a magnetic sample (with magnetization vectors also perpendicular to the surface),  and (ii) scattered into a ring of a radius reflecting the spatial correlation of magnetic domains (typically on 100 nm length scales, i.e., large in comparison with charge heterogeneity, $\sim$ 10 nm ), the overall scattering amplitude reduces to the magnetic contribution only \cite{Hannon98,Gutt10}. In the electric dipole approximation, it reads: 
%%%%%%%%
\begin{equation}
F^{\rm{magn}}= -\rm{i} \cdot  \left({\textbf{e}\times\textbf{e}^{\prime}}\right) \cdot \textbf{m}\,\,F^{\rm{m1}}.
\label{fmagnet}
\end{equation}
%%%%%%%%
The vectors $\textbf{e}$ and $\textbf{e}^{\prime}$ are polarization vectors of the incoming and scattered radiation, $\textbf{m}$ is the unit vector of the magnetization. The complex dipole-transition matrix element, $F^{\rm{m1}}$, describes the resonant magnetic scattering strength \cite{Brueckel}. It depends, among others, on the difference between the incoming photon energy and the resonant energy, and on the actual material magnetization, $M$. 

As the  resonant magnetic scattering strength,  $F^{\rm{m1}}$,  is proportional to the magnetization of the sample \cite{Mueller13,Andre21},  any changes of the magnetization within magnetic domains will be reflected by the change of the scattering signal. This, in particular,  implies that any demagnetization of the sample will cause a decrease of the magnetic signal. 

Let us emphasize that the modeling tool able to follow transient changes of magnetization has to take into account radiation damage processes in the sample induced by X-ray irradiation. The damage processes have, in general, two components: (i) electronic damage due to X-ray induced excitation and collisional relaxation of electrons, and (ii) structural damage resulting in atomic displacements. In this work, we report on a nanoscopic modeling tool, XSPIN, exploring only electronic damage. This restricts the applicability of our model to X-ray fluences below the structural damage threshold. However, this is the fluence regime of the strongest interest and applicational potential, as the demagnetization is then a reversible process and, after a certain time, the material recovers its equilibrium magnetization. If the structural damage fluence threshold is exceeded, the changes in the material become irreversible, and its magnetic properties can be  ultimately lost \cite{Gutt10,Wang2012,Mueller13}. 

With the XSPIN  tool, we analyze the results of a recent experiment \cite{Andre21} on resonant magnetic scattering with ultrashort XUV pulses (tuned to the M-edge of cobalt) from Co/Pt multilayer system with perpendicular magnetic anisotropy \cite{hellwig07}. The experiment was performed at the FERMI FEL facility. In particular, we  demonstrate that the processes of electronic excitation, relaxation and transport induced by XUV radiation predominantly affect the behaviour of the transient magnetization and hence the scattered magnetic signal.

%%%%%%%%%%%%%%%%%%%%%%
\section*{Results}
%%%%%%%%%%%%%%%%%%%%%%
%%%%%%%%%%%%%%%%%%%%%%
\subsection*{Modeling of X-ray induced processes in solid materials}
%%%%%%%%%%%%%%%%%%%%%%

The hybrid code XTANT \cite{Medvedev_njp,4open} (discussed in detail in the ''Methods'' section) was the base for the code XSPIN, which we have constructed and use in this study to follow X-ray induced magnetic transitions in solid materials. 

The XTANT code includes all predominant processes occurring in a solid material as a result of X-ray irradiation.  It is a hybrid simulation approach combining various modeling techniques. It enables  a treatment of large samples and highly excited electronic states (up to keV energy) which is not yet feasible with  fully {\it ab initio} approaches such as, e.g., those presented in \cite{werner19,sandholzer19}. The XSPIN code is  an extension of XTANT, which treats spin degrees of freedom in electronic subsystem.  The following paragraphs summarize the modeling framework of XSPIN. 

First, we assume that the incoming X-ray pulses are not intense enough to cause any atomic displacements in a magnetic material during the exposure. We neglect also eventual shifts of electronic levels due to high electron temperature. As the nuclei positions are fixed, we can use an  {\it ab initio} density of states (DOS) obtained for the material in equilibrium. For XSPIN simulations, it was calculated with the VASP (Vienna Ab initio Simulation Package) code which enables such high-precision DFT calculations for various materials~\cite{VASP,VASP1,VASP2,VASP3}. 

Second, in XSPIN we imply spin non-degeneracy to all electrons. Thus, each electron has its own spin, with two states: spin-up and spin-down. We exclude any spin precession, as it is negligible on the subpicosecond ($\sim$100 fs) timescales considered here. Consequently, the magnetic domains are static, and demagnetization only affects the magnitude of the magnetization in each individual domain.
The spin-up and spin-down valence electrons are initially distributed in the $3d$ band according to the total magnetic momentum of the material under thermal equilibrium. This scheme is similar to that used in the Stoner-Wolfarth model, describing a single magnetic domain \cite{stonerw47,tannous08}. 

\begin{figure}[h]
  \centering
\includegraphics[width=.8\textwidth]{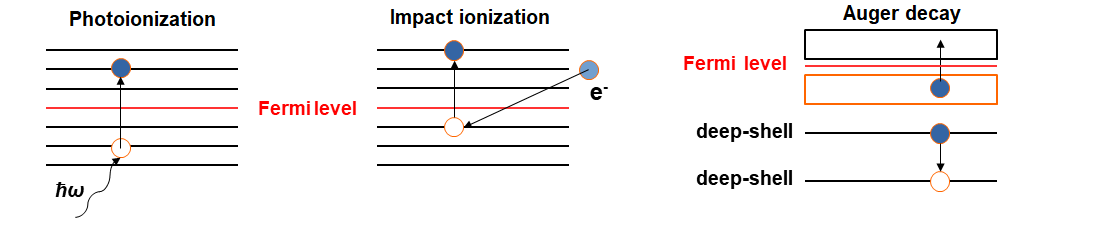} 
\caption{{\bf Excitation and relaxation processes treated by XSPIN code (schematically depicted).} (left) Photoionization, (middle) impact
ionization, and (right) Auger decay.}
  \label{Processes}
\end{figure}

Third, the code treats with different simulation techniques non-thermalized high-energy fraction (HEF) of electrons and thermally equilibrated low-energy fraction (LEF) of electrons, the latter involving electrons with energies below some specified energy cutoff (here, 15 eV).  The photoinduced and the subsequent electronic collisional processes involving the electrons within HEF are simulated with the classical Monte Carlo scheme adopted from the XTANT code~\cite{Medvedev_njp, 4open, Medvedev18}. Figure  \ref{Processes} shows schematically the electronic processes  considered, i.e., photoionization, impact ionization and Auger decay. After an X-ray pulse starts to interact with a solid material, electrons from spin-up and spin-down subsystems are released due to the photoabsorption process. The excitation probabilities take into account the actual electronic occupations in the respective bands. If the photon energy is sufficient to trigger an electronic excitation from a core shell, a spin-up or spin-down electron can be excited from the shell. After the photoabsorption, the energetic photoelectron joins the non-thermalized high-energy electron fraction, preserving its spin state. During the sequence of the following impact ionization events, the electron continuously loses its energy and may ultimately fall into the spin-up or the spin-down subsystem of the thermalized low-energy electron fraction -- depending on its spin state. The HEF electrons may excite further electrons, with the same or an opposite spin. The probability of such excitation depends on the actual occupations of the spin-up and spin-down electron levels in the LEF and in the core shells.

Core holes  relax via Auger decay. A band electron with the same spin fills the hole, while the Auger electron is chosen randomly, according to the actual distribution of spin-up and spin-down electrons. The cross sections for photoionizations are taken from the EPDL database~\cite{EPDL} and for impact and Auger ionizations from the EADL database~\cite{EADL}. The core ionizaton potentials are taken from the X-ray Data Booklet~\cite{XDB}. 

All low energy electrons from the LEF, both within spin-up and spin-down subsystems, are assumed to stay in a {\it common} local thermal equlibrium. Therefore, at each time step {\it all} the electrons are instantly thermalized to follow a Fermi-Dirac distribution. Note that the intraband collisions leading to the thermalization in the entire low-energy electronic subsystem must then also include spin-flip collisions. In other words, the spin redistribution in our model occurs through thermal collisions. Let us emphasize that XSPIN does not trace the overall angular momentum of the system.

At each time step,  the actual number of electrons and the actual electron energy stored in both spin-up and spin-down subsystems of the electronic LEF ($N_{\rm{e}}^{\rm{low}}$ and $E_{\rm{e}}^{\rm{low}}$ respectively) are followed. Knowing them, a common temperature, $T_{\rm{e}}$, and a common chemical potential of electrons,  $\mu$, can be calculated  by solving the equations:
%%%%%%%%
\begin{eqnarray}
N_{\rm{e}}^{\rm{low}} = \sum_{\sigma}\sum_{E_{{\rm{min}},\sigma}}^{E_{{\rm{max}},\sigma}}\{1 + {\rm{exp}}[(E_{i,\sigma} - \mu)/(k_{\rm{B}} T_{\rm{e}})]\}^{-1}, \
\nonumber \\
E_{\rm{e}}^{\rm{low}} = \sum_{\sigma}\sum_{E_{{\rm{min}},\sigma}}^{E_{{\rm{max}},\sigma}} E_{i,\sigma} \{1 + {\rm{exp}}[(E_{i,\sigma} - \mu)/(k_{\rm{B}} T_{\rm{e}})]\}^{-1}, 
\label{bisection}
\end{eqnarray}
%%%%%%%%
similarly as it was done in the code XTANT,  therein with spin degeneracy ~\cite{Medvedev_njp}. The energy $E_{i,\sigma}$ is the energy of the $i$-th level for spin $\sigma$ = $\uparrow$, $\downarrow$; $E_{{\rm{min}},\sigma}$ and $E_{{\rm{max}},\sigma}$ are  the minimal and maximal (cut off) electronic energies in the band $\sigma$ respectively, and $k_{\rm{B}}$ is the Boltzmann constant.
The energy levels in the low-energy electron fraction are determined from the total spin-polarized density of states $D_{\sigma}(\epsilon)$ for fcc Co, calculated with  the code VASP. The energy $E_{i,\sigma}$ of $i$-th level for spin-$\sigma$ electrons is then calculated from the equation: $i = \int_{-\infty}^{E_{i,\sigma}} d \epsilon \, D_{\sigma}(\epsilon)$. Also, the number of energy levels is determined by this equation. For example, for 64 atoms the number of electron levels between the bottom of d-band and the cut-off energy of 15 eV is 481 for spin-up electrons, and 474 for spin-down electrons.
%%%%%%%%%%%%%%%%%%%%%%

Below we list further features of the XSPIN model:

(1) We assume that the photons scattered due to the resonant magnetic scattering process do not induce further magnetic scattering. This assumption is justified by the very small cross section for the resonant magnetic scattering in comparison to the photoabsorption cross section \cite{xatom}.

(2) We also assume that the X-ray fluences applied do not cause a significant structural damage to the material during or shortly after the XUV pulse, i.e., on $\sim$100 fs timescales.  We give the justification below.
A rigorous definition of structural damage threshold is difficult at the 100 fs timescale considered. The usual measure for a damage threshold in a metal is the threshold dose for its thermal melting. This dose for cobalt is estimated as ~0.54 eV/atom. However, the thermal melting would require picosecond(s) to be completed. This time is needed for a transfer of a sufficient amount of energy from the electronic system to the lattice. At 100 fs timescale, we can only use this threshold dose as an indicator when structural modifications can start to play some role. 
 
Therefore, in our model, we can assume that the atomic positions are fixed, i.e., the atoms do not change their positions during the simulations. This is because the timescale of atomic displacements during the structural transformation is then longer than the femtosecond pulse duration, see, e.g., \cite{fermigraph,inoue21,melting21}.We also neglect eventual shifts of electronic levels due to high electron temperature. These both assumptions guarantee a reasonable modeling accuracy even for the doses a few times higher than 0.54 eV/atom, on 100 fs timescales.  However, at higher absorbed X-ray doses or if the model should be applied at picosecond timescales (e.g., in order to follow the recovery of the magnetization), the possible atomic relocations should be taken into account. Such an extension of XSPIN is possible but it would require a significant modification of the anyway complex code, with much effort to be invested. Still, we plan this effort in future.

(3) We assume that all band electrons (both from the spin-up and from the spin-down fractions) undergo instantaneous thermalization at each time step. The intraband collisions, which lead to the electron thermalization, also include spin-flip collisional processes between spin-up and spin-down electrons (cf. \cite{diffspin11}). In such a way, the spin-flip processes are implicitly included in our model.  Electron--ion coupling is neglected here, due to ultrashort timescales considered.

Note that the assumption of the instantaneous electron thermalization limits the applicability of the XSPIN to model X-ray irradiation with X-ray pulses of duration longer than the timescale of electronic thermalization. We have performed dedicated simulations with the XCASCADE(3D) code \cite{Lipp17} to investigate the timescale of electron cascading process in Co and Pt, which is comparable to the timescale of electron thermalization.  The calculations show that a photon of energy $\sim$ 61.1 eV (as used in the experiment)  creates on average $3.81$ electrons in Co and $5.22$ electrons in Pt within $0.2$--$0.4$ fs, both through the excitations from valence band and from the uppermost core levels. This indicates that the XSPIN model should not be applied for subfemtosecond X-ray pulses.

(4)  Interactions between magnetic domains in $(X,Y)$ plane are not included, consistently with the Stoner-Wolfarth model framework of a single magnetic domain \cite{stonerw47,tannous08}, used here. Results from a simplistic model with periodic domains (not shown) indicate  that the details on domain structure in $(X,Y)$ plane should not significantly affect our results on 100 fs timescales.

(5) X-ray pulses from FERMI facility have a high degree of coherence, as documented in Ref. \cite{Flavio13}. Therefore, the total signal scattered from the multilayer sample is calculated as a coherent superposition of the contributions from individual layers.

(6) For the XSPIN analysis, we used average fluence values estimated by the experiment \cite{Andre21}.  They were estimated, knowing the beam energy focused into a FWHM focal spot. We  assumed that the spatial profile of X-ray pulse in our simulations was flat-top, with an average fluence.  No volume integration of the signal in the $(X,Y)$ plane was performed. For a meaningful volume integration, we would need much more precise information on the spatial pulse profile than provided by the experiment, in particular, the information on the pulse wings shape.

(7) We included the effect of interlayer electron transport in our predictions. The significant role of electron transport in demagnetization processes was indicated in earlier works on diffusive spin currents, e.g., \cite{Pfau12,diffspin10,diffspin12}. In the 100 fs time regime, considered in this work (with the XUV pulse of duration 70 fs, acting both as a pump and as a probe), only the start phase of the spin currents, i.e., the ballistic transport regime, can be treated. Treatment of long-range electron transport is not necessary because such transport does not have enough time to develop. We then only focus on fast collisional processes influencing the electron distribution within the magnetically sensitive regime of the $3d$ band. X-ray photoabsorption processes cause the emission of electrons, both through direct photoionization as well as through the $3p$ Auger processes. These ballistic electrons can excite further electrons in collisional processes. The resulting electron cascades then spread in the material. In our multilayer sample, the cascade electrons can also enter the neighbouring layers. As mentioned in  (3),  dedicated simulations with the XCASCADE(3D) code \cite{Lipp17} predicted that a photon of energy $\sim$ 61.1 eV (used in the experiment)  created on average $3.81$ electrons in Co and $5.22$ electrons in Pt within $0.2$--$0.4$ fs, i.e., almost instantaneously, both through the excitations from valence band and from the uppermost core levels.  The averaging has been performed over $30000$ Monte-Carlo cascade realizations. The electron ranges \cite{b4c}, i.e., the maximal distances traveled by electrons released in a single photoabsorption event until their energy decreased below ionization threshold, were $1.49$ nm and $10.51$ nm for Co and Pt respectively. This clearly indicates that interlayer electron transport cannot be neglected in our multilayer sample, where layer thicknesses are only: $d_{\rm{Co}}   = 0.8$ nm and $d_{\rm{Pt}}   = 1.4$ nm.  After the electron cascading stops, modifications within the magnetically sensitive regime of $3d$ band through collisional ionization processes stop as well. Low energy electrons  propagate further through the material in a diffusive transport. As the energy of these electrons is located within the magnetically sensitive regime of $3d$ band, their diffusive transport throughout the sample, followed by interactions with local $3d$ electrons, can further modify magnetic properties of the system. However, it occurs on much longer timescales, as indicated by earlier works on diffusive spin currents, e.g., \cite{diffspin10,diffspin12,Pfau12}.

The interlayer electron transport was modeled in the following way in the XSPIN code. First, the number of additional electrons in each Co layer which arrived from other layers was estimated, knowing the distribution of the absorbed photons and the electron range in Co and Pt materials.  Repeated calculations were then performed with XSPIN, assuming a higher (effective) X-ray pulse fluence, such that would lead to the production of the increased number of electrons (including additional electrons originating from interlayer transport). The XSPIN results on the resonant magnetic scattering signal from a multilayer system tested in the experiment \cite{Andre21}, shown later in the paper,  were calculated, taking the interlayer electron transport into account. 

(8) XSPIN simulations were performed for the supercell containing 64 Co atoms. As we consider fluences low enough not to cause atomic relocations, such number of atoms is sufficient to get a statistically reliable results. This expectation was confirmed by the preceding convergence tests of our results in respect to the size of the supercell (not shown).

\begin{figure}[h]
	\centering
     \includegraphics[width=0.7\textwidth]{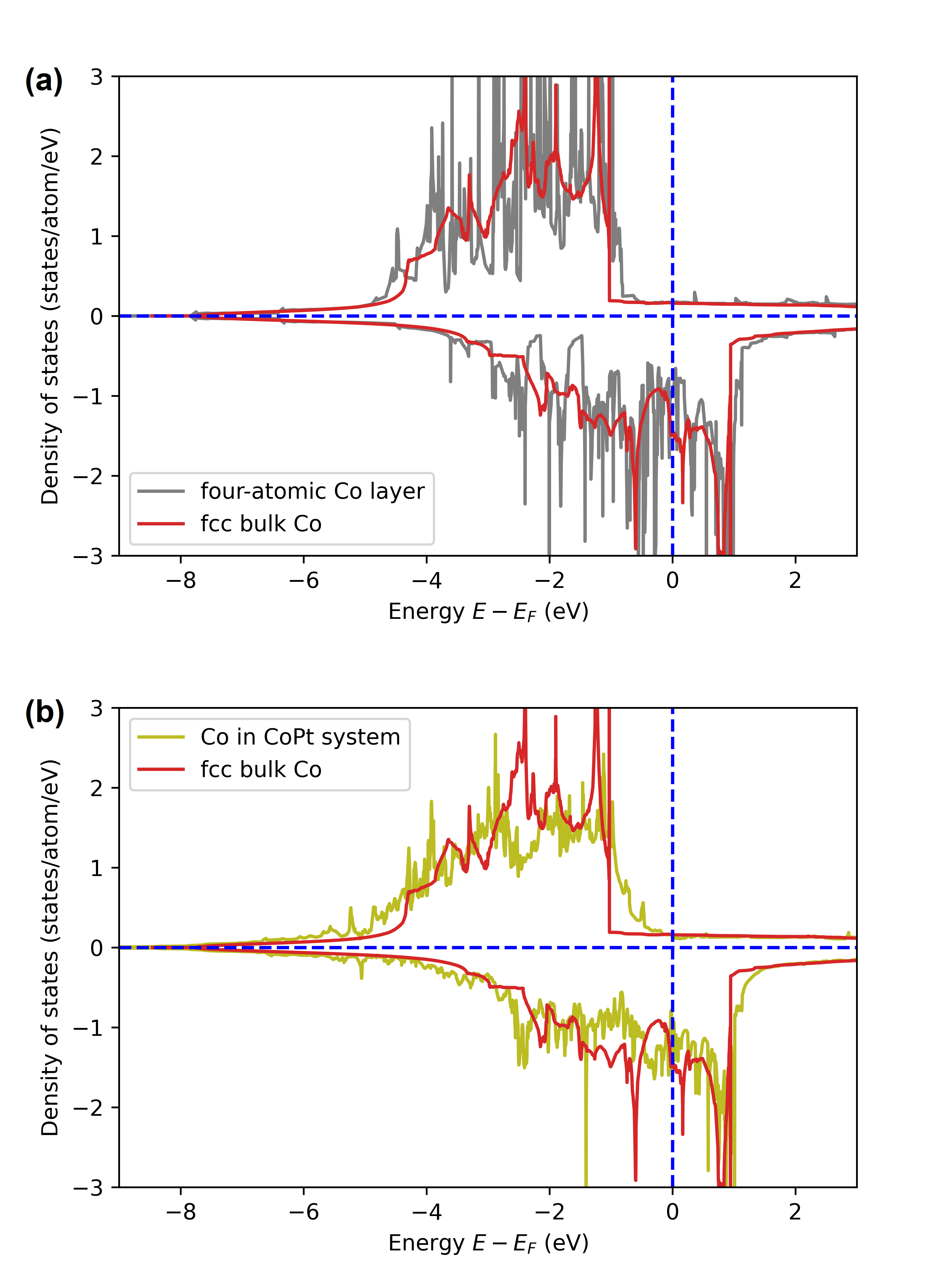}
    \caption{{\bf Calculated densities of states.} Comparison between: {\bf a} density of states calculated for 4-atomic-layer Co structure and the density of states calculated for bulk Co, and {\bf b} partial density of states for Co atoms extracted from Co-Pt multilayer structure (4 atomic layers of Co followed by 6 atomic layers of Pt) and the density of states calculated for bulk Co. All calculations were performed with the VASP code~\cite{VASP,VASP1,VASP2,VASP3}.}
    \label{DOSs}
\end{figure}

(9) Accuracy of DOS calculations performed for Co/Pt multilayer system: In Figs.\ \ref{DOSs}a and \ref{DOSs}b, the comparison is shown between: (i) the density of states  calculated for a 4-atomic-layer Co structure and the density of states calculated for bulk Co, and (ii) between the partial density of states extracted for Co atoms from the Co-Pt multilayer structure (4 atomic layers of Co followed by 6 atomic layers of Pt) and the density of states calculated for bulk Co atoms. The presented results clearly indicate that there are no significant differences between the calculated density of states in all considered cases. In particular, the overlap between Co and Pt electronic orbitals is of minor importance. Moreover, the densities of state for the 4-atom Co layer in the vacuum and for the bulk Co system look similar. Thus, the usage of the density of states obtained for bulk Co for the parametrization of low-energy electronic levels in our code XSPIN seems well justified.

\begin{figure}[h]
	\centering
    \includegraphics[width=0.7\textwidth]{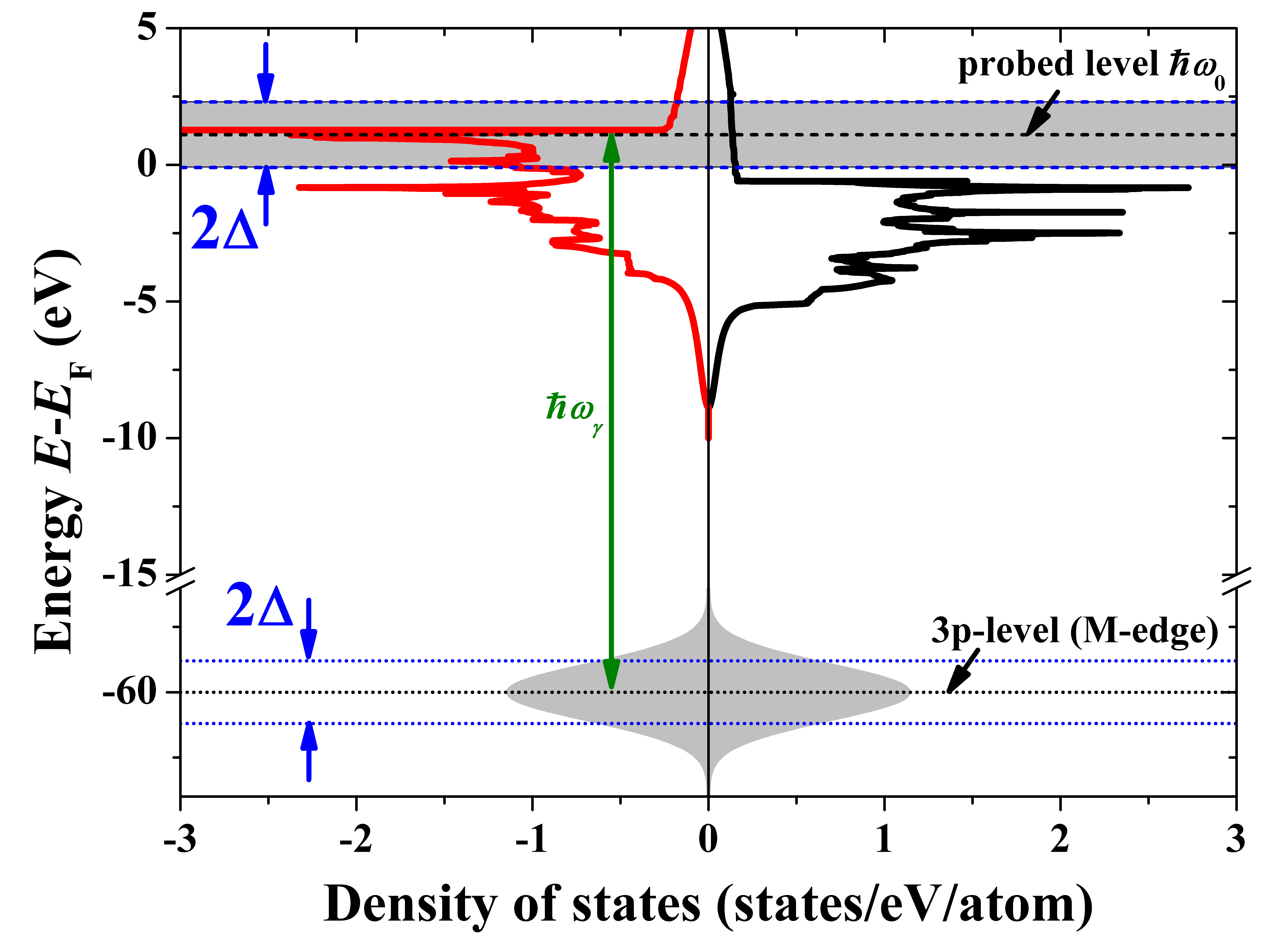}
    \caption{{\bf Calculated density of states for equilibrium fcc cobalt}, with schematic indication of the $3p$ band of cobalt, and of the probed region in its $3d$ band. The width of the $3p$ band is $2\Delta$.}
    \label{DOS}
\end{figure}

%%%%%%%%
Now we will  describe in detail how magnetic signal from X-ray irradiated Co layer is constructed in XSPIN. X-ray magnetic dichroism arises from a directional spin alignment and the spin-orbit coupling, and results in different X-ray absorption of left and right circularly polarized light at the absorption edges of ferromagnetic materials \cite{Stoehr06}. The absorption spectra reflect the actual positions of electronic energy levels and the actual occupations of the resonant electronic states.  Let us consider a magnetic scattering signal from an X-ray irradiated Co layer. Incoming X-ray photons of energy $\hbar\omega_{\gamma}$ can then excite electrons  from the $3p$ band to the $3d$ band (Fig.~\ref{DOS}). The region in the $3d$ band to which the electrons can be excited from the $3p$ band extends from  $\hbar\omega_0 - \Delta$ to  $\hbar\omega_0 + \Delta$, where $\hbar\omega_0$ is the difference between the photon energy and the position of M-edge:
%%%%%%%%
\begin{equation}
\hbar\omega_0 = \hbar\omega_{\gamma} - E_{\rm{edge}},
\label{probed_level}
\end{equation}
%%%%%%%%
with $E_{\rm{edge}}$ = 60 eV for M-edge of Co. Here, $2\cdot \Delta$ is the $3p$ band width, which determines the number of states probed in the $3d$ band.
%%%%%%%%
The magnetization  is  proportional to the disparity between electronic populations at the resonant states in spin-up and spin-down subsystems:
%%%%%%%%
\begin{equation}
M \propto \sum_{\hbar\omega_0-\Delta}^{\hbar\omega_0+\Delta}[N_{\uparrow}^{\rm{hole}}(E_{i,\uparrow}) - N_{\downarrow}^{\rm{hole}}(E_{i,\downarrow})],
\label{magn-signal}
\end{equation}
%%%%%%%%
where $N_{\sigma}^{\rm{hole}}(E_{i,\sigma}) = 1 - N_{\rm{e},\sigma}^{\rm{low}}(E_{i,\sigma})$ denotes the number of empty states at $E_{i,\sigma}$ level. The coefficient $N_{\rm{e},\sigma}^{\rm{low}}(E_{i,\sigma}) = \{1 + {\rm{exp}}\{[(E_{i,\sigma} - \mu)/(k_{\rm{B}} T_{\rm{e}})]\}^{-1}$ defines the electronic occupation of the level $E_{i,\sigma}$ (assumed to be a Fermi-Dirac occupation at all times). 

The XSPIN code calculates transient changes of $N_{\sigma}^{\rm{hole}}(E_{i,\sigma})$ in response to a specific X-ray pulse for the probed energy levels within the Co $3d$ band (i.e., within the interval $\pm \Delta$ around the probed level $\hbar\omega_0$). The transient magnetization of the system can then be calculated from Eq.\ (\ref{magn-signal}). 

\begin{figure}[h]
  \centering
\includegraphics[width=.7\textwidth]{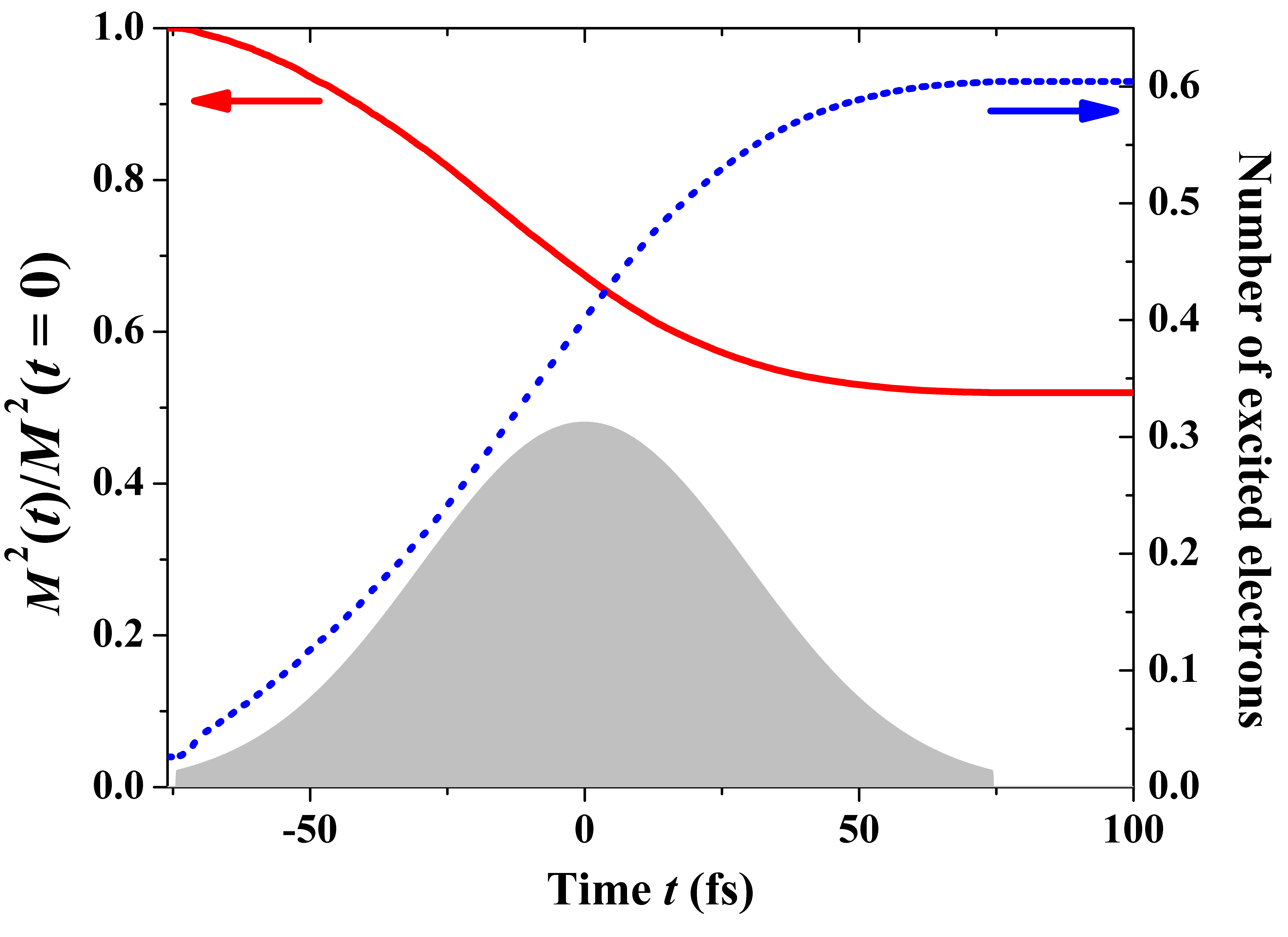}
 \caption{{\bf Normalized magnetization in a single Co layer as a function of time} (red solid line), and {\bf the transient number of excited electrons}, i.e., the electrons with energy above the Fermi level (blue dashed line), calculated with XSPIN for XUV pulse fluence,  $F= 13$ mJ/cm$^2$, corresponding to average absorbed dose of 0.93 eV/atom. Photon energy was 61.1 eV, as in the experiment \cite{Andre21}. The temporal pulse profile is schematically shown.}
\label{magnet-M}
\end{figure}

Figure \ref{magnet-M} shows an example of a typical shape of the demagnetization curve,   $|M(t)|^2$, normalized to its initial value, $|M(t=0)|^2$, obtained for a single Co layer. The temporal shape of the X-ray pulse was Gaussian, with the full width at half maximum (FWHM) of 70 fs. The  pulse fluence was 13 mJ/cm$^2$, corresponding to the average absorbed dose in the material of 0.93 eV/atom. The assumed thickness of the Co layer was 0.8 nm, i.e., much less than the photon attenuation length in Co for a $\sim$60 eV photon. This  ensured a uniform distribution of absorbed energy within the Co layer. The decrease of the $|M(t)|^2$ curve follows the increase of the number of excited electrons (i.e., the electrons with energy above the Fermi level) in the sample, also depicted in  Fig. \ref{magnet-M}. When the electron cascading saturates, the value of $|M(t)|^2$ stabilizes, here within $\sim$70 fs after the pulse maximum. 

The transient intensity of the resonant magnetic scattering signal (per unit surface), $I^{\rm{magn}}(t)$,  (cf. \cite{Gutt10,Andre21,Stoehr06}) is:
%%%%%%%%
\begin{equation}
I^{\rm{magn}}(t) \propto  I(t)  \cdot |F^{\rm{magn}}(t)|^2
\end{equation}
%%%%%%%%
where $I(t)$ is the incoming X-ray intensity, and  $F^{\rm{magn}}(t)$ is the instantaneous amplitude for the resonant magnetic scattering taken from Eq.\ (\ref{fmagnet}). If we separate out the magnetization from the the dipole-transition matrix element in $F^{\rm{magn}}(t)$ (see, e.g., \cite{Stoehr06}), assuming that energy level shifts and stimulated emission processes are only of minor importance -- which is the case here -- the magnetic scattering signal can be rewritten as:
%%%%%%%%
\begin{equation}
I^{\rm{magn}}(t) \propto I(t)  |M(t)|^2,
\label{mstrength}
\end{equation}
%%%%%%%%
where $M(t)$ is the transient magnetization.  The time-integrated intensity, $I^{\rm{magn}}(t)$, yields the experimental observable, magnetic scattering efficiency, $S(F)$:
%%%%%%%%
\begin{equation}
S(F) = P \cdot \int  dt\, I(t)  |M(t)|^2,
\label{scatt-eff}
\end{equation}
%%%%%%%%%
where $F=\int \, dt\,  I(t)$ is the pulse fluence. The proportionality factor, $P$,  in Eq.\ (\ref{scatt-eff}) depends both on the material properties and on  the X-ray beam parameters (e.g., polarization) \cite{Andre21}. However, it does not depend on the X-ray pulse fluence.

The formalism presented above works only for the case when a single XUV pulse serves both as a pump and a probe, and the time-integrated mSAXS signal is recorded. Simulation of measurements with separate pump and probe pulses (e.g., \cite{hennes21}) would require a dedicated treatment with a respective modification of the Eq.\ (\ref{scatt-eff}), taking into account the actual time delay between the probing time and the response of the magnetic system to the pump pulse (e.g., \cite{carva09}).
%%%%%%%%%%%%%%%%%%%%%%%%%%%%%%%
\subsection*{Magnetic signal recorded in mSAXS experiments}
\begin{figure}[h]
	\centering	
     \includegraphics[width=0.7\textwidth]{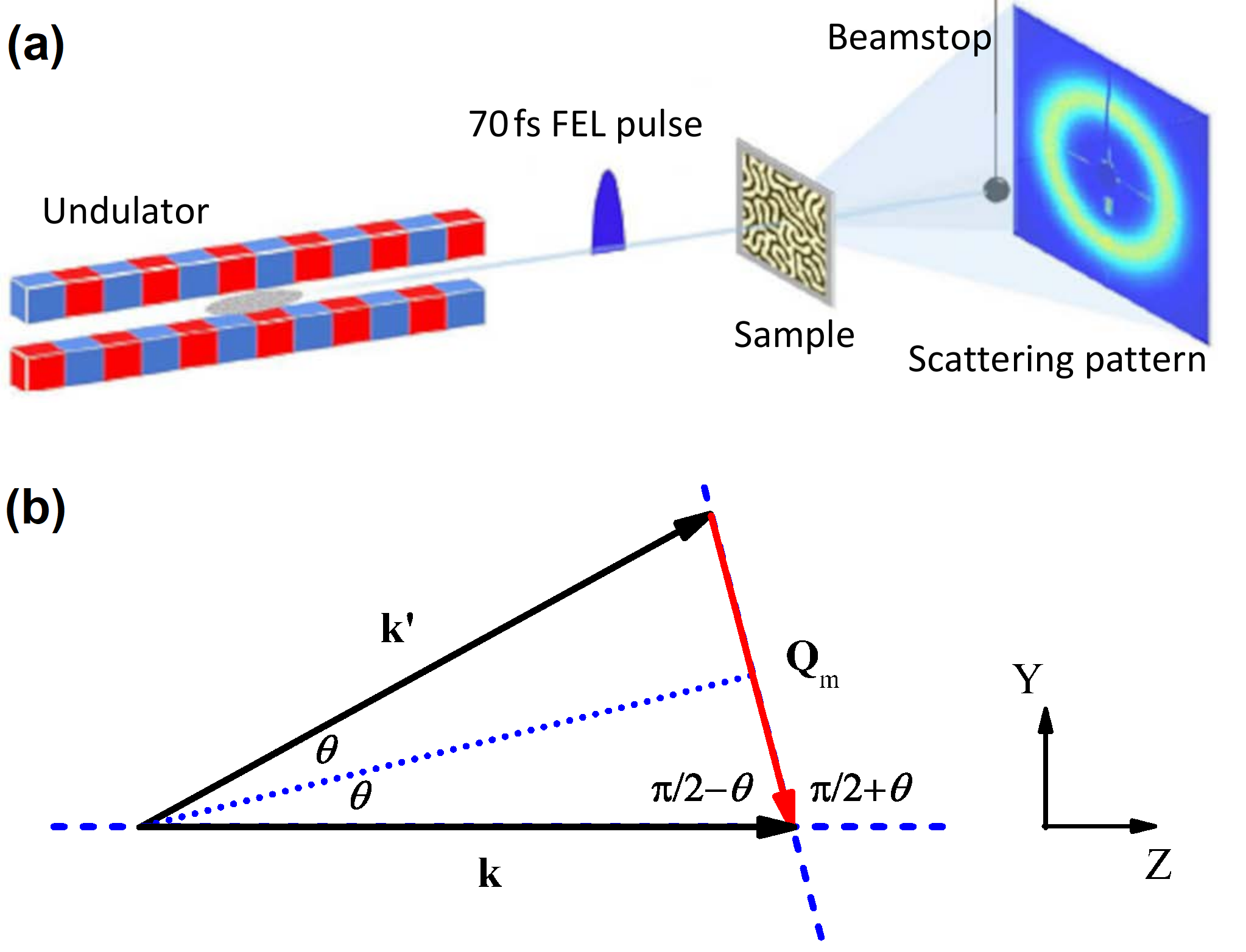}
\caption{{\bf Principle of mSAXS measurement.} {\bf a} scheme of the mSAXS setup, and {\bf b} relation between the scattering vector $\textbf{Q}_{\rm{m}}$ and  the incoming and scattered wave vectors $\textbf{k}$ and $\textbf{k}^{\prime}$ during magnetic scattering.}
    \label{Qgeometry}
\end{figure}

Experiments investigating X-ray induced demagnetization use multilayer systems in order to strengthen the overall magnetic scattering signal (which then becomes a sum of contributions from individual layers), and to tune the magnetic domain size, see, e.g., \cite{Bran10}. The multilayer systems are composed of ferromagnetic and paramagnetic materials, e.g., Co/Pt \cite{Gutt10,Bran10,Andre21} or Co/Pd \cite{Wu,Wang2012,Schneider2020}. In multilayer samples of such composition, magnetic maze domains are formed, with magnetization perpendicular to the layer surface and alternating up and down \cite{Andre21}. In an mSAXS experiment, coherent  X rays arrive with normal incidence at the top layer of the multilayer system and propagate through it (Fig.\ \ref{Qgeometry}a). Two processes can then occur: (i) photoabsorption, and (ii) coherent scattering including resonant magnetic scattering if the radiation is tuned to the absorption edge. Magnetically scattered photons are recorded at the CCD detector. They form a scattering ring which radius reflects the spatial correlation of magnetic domains, being twice the domain size, $\zeta=2 {\rm{\pi}}/Q_{\rm{m}}$ (Fig.\ \ref{Qgeometry}b), where $Q_{\rm{m}}$ is the length of the scattering vector $\textbf{Q}_{\rm{m}}$, $Q_{\rm{m}} = \frac{4 {\rm{\pi}} }{\lambda} \sin \theta$, with $\lambda$ being the wavelength of the incoming radiation, and $2\cdot  \theta$  being the scattering angle.  The total intensity of the ring reflects time-integrated scattering efficiency of magnetic domains \cite{Pfau12}.

Figure \ref{Co-Pt} shows schematically the multilayer system studied in the experiment by Kobs et al. \cite{Andre21} for which we will later present the corresponding XSPIN predictions. The FEL beam was first impinging at normal incidence on the $\rm{Si_3N_4}$ membrane (not shown) and then entered the top platinum layer, $\rm{Pt}_{\rm{in}}$. The spatial correlation, $\zeta$, of the maze domains in the Co layers was of the order of 180 nm (corresponding to the peak scattering vector $Q_{m}=0.036$ nm$^{-1}$), and lead to a pronounced mSAXS  signal. The experiment used incoming photons of energy $\sim 61.1$ eV. Pulse fluences on the top Pt layer, $F_{\rm{Pt,in}}$, were between 0.3 and 45 mJ/cm$^2$. The temporal shape of the XUV pulse was Gaussian, with full width at half maximum (FWHM) of 70 fs. 
%%%%%%%%%%%%%%%%%%%%%%%%%
\subsection*{Theory predictions for mSAXS signal}
Our goal is now to validate the XSPIN model predictions. For this purpose, we used the already existing experimental data from the mSAXS experiment \cite{Andre21} performed with photon energies tuned to the Co M-edge, which used Co/Pt multilayers.

\begin{figure}[h]
  \centering
   \includegraphics[width=0.8\textwidth]{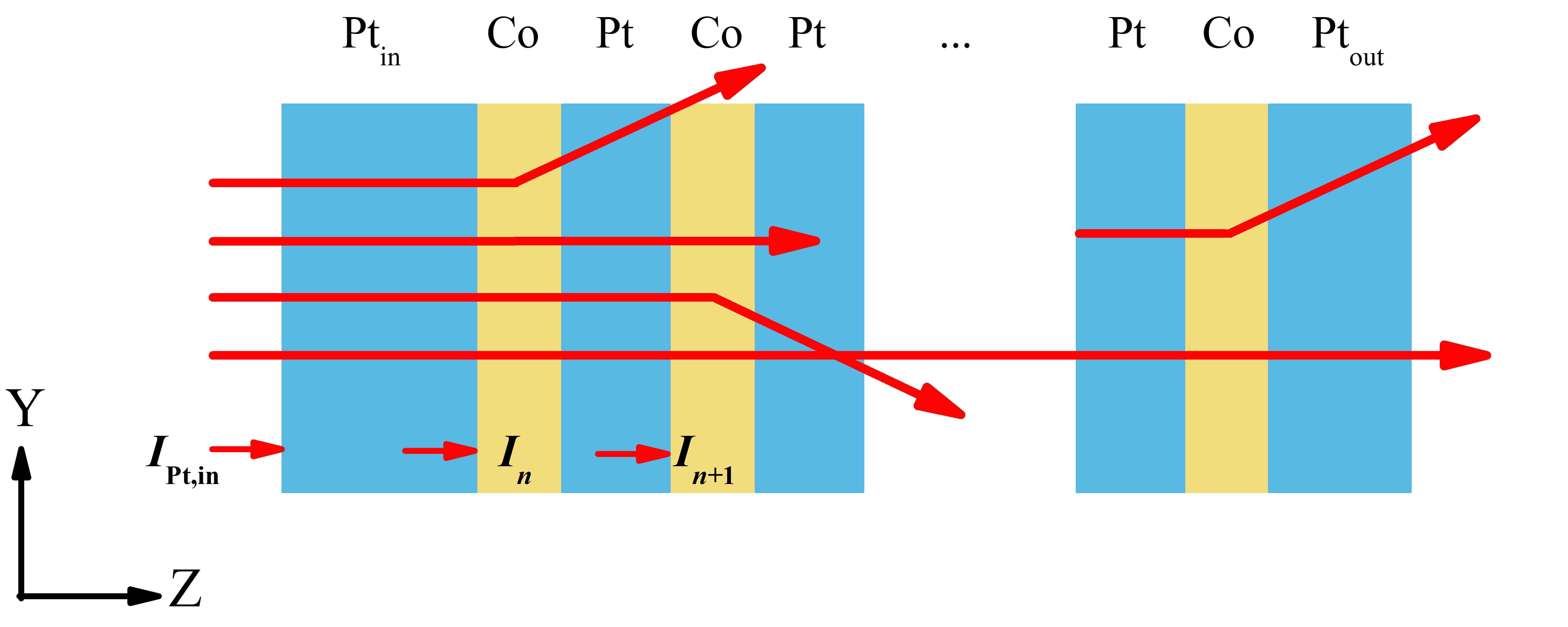}
    \caption{{\bf Schematic view of the Co/Pt multilayer system with incoming, scattered and absorbed radiation.} The system  $\rm{ Pt(5.0 nm)/[Co(0.8nm)/Pt(1.4 nm)]_{16}/Pt(0.6 nm) } $ used in \cite{Andre21} consists of 5 nm thick Pt$_{\rm{in}}$ layer, 16 layers of Co (each 0.8 nm thick), alternating with 15 layers of Pt (each 1.4 nm thick), and 2.0 nm thick Pt$_{\rm{out}}$ layer.  The actual sample is deposited on a 50 nm thick $ \rm{Si_3N_4}$ membrane acting as structural support placed before the Pt$_{\rm{in}}$ layer (not shown here). The absorption of incoming radiation in this layer has been taken into account in our analysis.}
  \label{Co-Pt}
\end{figure}

In order to describe the response of the multilayer system  (Fig.\ \ref{Co-Pt}) to X-ray/XUV irradiation, one has to analyse propagation of the radiation through the system. Let us first note that any reflection on Co or Pt layers can be neglected as the reflectivity coefficients for Co and Pt layers at 61.1 eV photon energy are of the order of $\sim 10^{-2}$--$10^{-3}$. The change of incoming X-ray intensity after passing through a layer of a material can then be expressed as:
%%%%%%%%
\begin{equation}
{ {dI}\over{dz}}= -\alpha_{\rm{phot}} \cdot \, I,
\label{photoabs}
\end{equation}
%%%%%
\noindent where $\alpha_{\rm{phot}}$ is the photoabsorption coefficient, equal to the inverse of the photon attenuation length \cite{b4c}. The solution of this Beer-Lambert-type equation is well-known (see, e.g., \cite{b4c,Scherz}). According to it, X-ray pulse intensity changes as:
%%%%%%
\begin{equation}
I = I_0 \cdot e^{-d_{\rm{Co}}/\lambda_{\rm{att,Co}}},
\label{imagnet}
\end{equation}
%%%%%%
after passing through a Co layer of a thickness $d_{\rm{Co}}$, where  $\lambda_{\rm{att,Co}}$ is the photon attenuation length in cobalt. In our multilayer system (Fig.\ \ref{Co-Pt}), this implies a recursive relation between the X-ray intensities in two consecutive Co layers: the $n$th layer and the $(n+1)$th layer of the same thickness $d_{\rm{Co}}$ at a time instant $t$,
%%%%%%%
\begin{equation}
I_{n+1}(t)=  I_{n}(t)\cdot a_{\rm{Co}}\cdot a_{\rm{Pt}},
\label{twostars}
\end{equation}
%%%%%%%
% FROM Beata's code: final-el-dist.f; Photon energy 61.1 eV
%%%%%%%
where $\lambda_{\rm{att,Pt}}$ is the photon attenuation length in platinum, and $a_{\rm{Co(Pt)}}$ are material attenuation coefficients in Co or Pt, defined as $a_{\rm{Co(Pt)}}=e^{ - d_{\rm{Co(Pt)}}/\lambda_{\rm{att,Co(Pt)}} }$. For 61.1 eV photons, $\lambda_{\rm{att,Co}}\sim 9.20 $ nm and $\lambda_{\rm{att,Pt}}\sim 9.13$ nm respectively. They are $\sim$ 4 times shorter than the overall thickness of the multilayer system ($40.8$ nm), i.e., the pulse intensity attenuation has to be taken into account. The initial condition for Eq.\ (\ref{twostars}) is: $I_1(t)=I_{\rm{Pt,in}}(t)\cdot a_{\rm{Pt,in}}$. Here we neglected any intensity attenuation due to resonant magnetic scattering as the corresponding cross section is much smaller  than the photoabsorption cross section \cite{xatom}. 

The time-integrated scattered signal emitted into the magnetic ring, $Q_{\rm{m}}$, is then a coherent sum of contributions from different Co layers within the multilayer system:
%%%%%%%%
\begin{equation}
S(F_{\rm{Pt,in}};Q_{\rm{m}}) = P \cdot \, \int\, dt \, \left|
\sum_{n=1}^{N_{\rm{Co}}} \, \sqrt{ I_n(t)} \cdot M(t)  \cdot 
\sqrt{ a_{\rm{Pt}}^{N_{\rm{Co}}-n} \cdot a_{\rm{Co}}^{N_{\rm{Co}}-n}\cdot a_{\rm{Pt, out}} }\cdot
e^{-\rm{i} \,\rm{Q}_{\rm{m}}\cdot \rm{R}_n} 
\right|^2,
\label{sqeq}
\end{equation}
%%%%%%%%
\noindent where $I_n(t)$ also contains attenuation coefficients (see Eq.\ (\ref{twostars})). It can be shown that the overall product of the attenuation coefficients in Eq.\ (\ref{sqeq}) is the same for each layer,  i.e., the total signal sums the contributions from different layers with the same attenuation weight. Note also that Eq.\ (\ref{sqeq}) accounts for the fact that during the passage of 61.1 eV photons through Pt layers only photoabsorption processes and no resonant magnetic scattering occur.  To justify the latter assumption, we checked that platinum has absorption edges at 54 eV and 66 eV. However, as Fig. 2b in Ref. \cite{Willems17} shows, the contribution of Pt magnetic resonances to the overall resonant magnetic scattering signal from Co and Pt at 61.1 eV is subleading, when compared with the contribution of Co. To illustrate, platinum peak heights at 54 eV and 66 eV are $\sim$0.12 and $\sim$0.15 respectively, and Co peak height is $\sim$0.75 at 61.1 eV,  see Fig. 2b in  \cite{Willems17}. Therefore, in the framework of our model, we neglected Pt contribution to the resonant scattering signal at 61.1 eV.
%%%%%%%%%%%%%%%%%%%%%%%%%%%%%%%%%%
\subsection*{Comparison of XSPIN  predictions to experimental data.}

\begin{figure}[h]
  \centering
 \includegraphics[width=.7\textwidth]{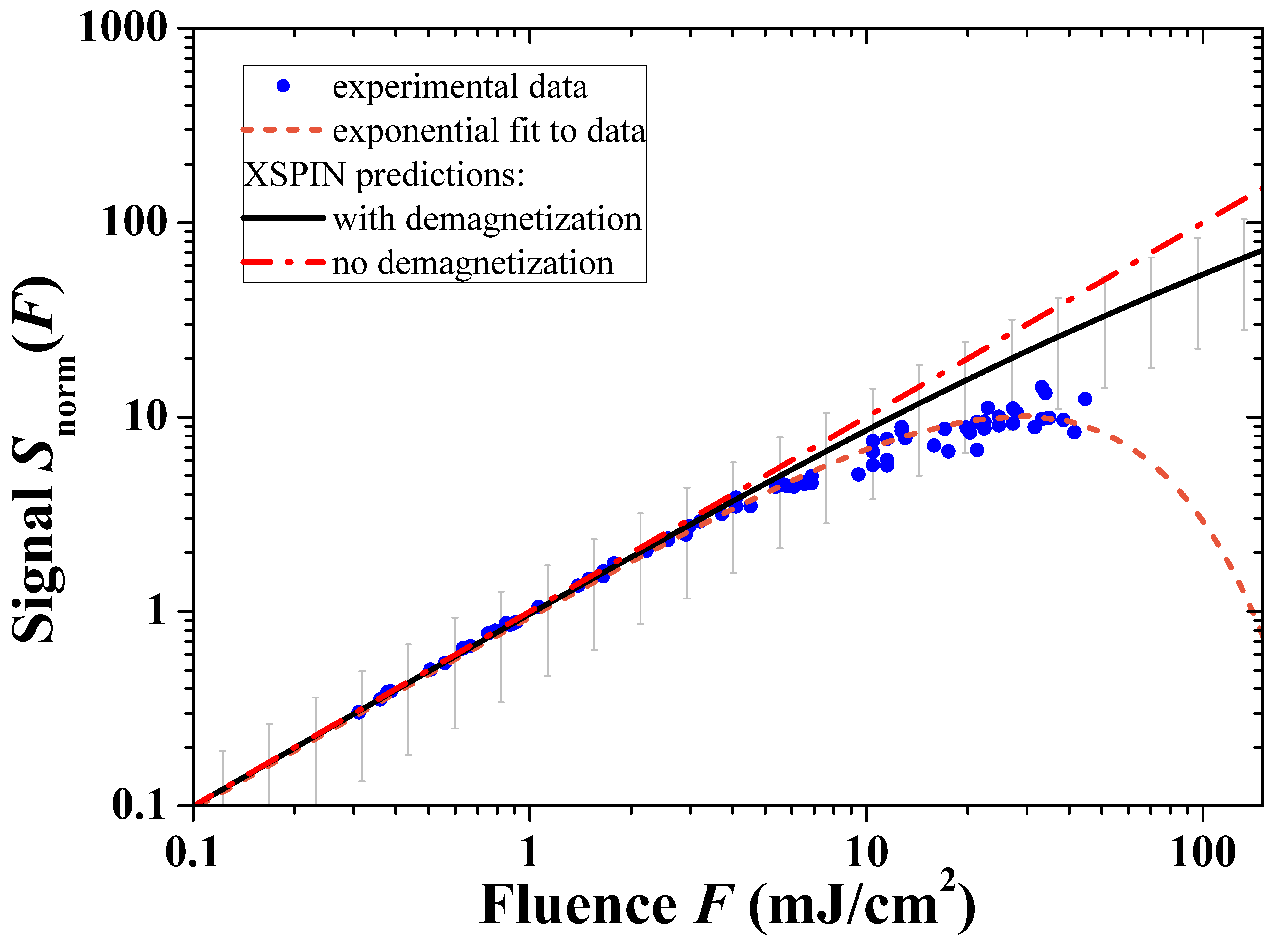}
 \caption{{\bf Normalized resonant magnetic scattering signal, $S_{\rm{norm}}(F)$ at Co M-edge} (in fluence units) shown as a function of the incoming fluence for Co/Pt multilayer system tested in \cite{Andre21}. Experimental data from \cite{Andre21} (blue points) are shown with the exponential fitting curve to the experimental data (orange dashed line), and with the theoretical predictions of the XSPIN code for $\Delta=1.2$ eV, taking interlayer electron transport into account. Predictions including the demagnetization (black solid line), and predictions assuming no demagnetization, i.e., $M(t)=M(0)$ (red dash-dotted line) are shown for comparison. The error bars give an estimate of the uncertainty of the applied signal scaling.}
\label{sqfig}
\end{figure}

We have calculated the magnetic scattering signal numerically, using XSPIN results for time-dependent magnetization obtained for various (attenuated) X-ray fluences in each Co layer of the multilayer system. Interlayer electron transport was also taken into account (for details, see ''Further features of XSPIN model'' section).  In what follows, we will use a simplified notation: $F\equiv F_{\rm{Pt,in}}$ and  $S(F) \equiv S(F,Q_{\rm{m}})$. Figure \ref{sqfig} shows the result on the normalized magnetic scattering signal, defined as: $S_{\rm{norm}}(F)=S(F) \cdot F_0/S(F_0)$, for $\Delta=1.2$ eV.  This value of $\Delta$ corresponds to the half of the FWHM of the Co M-edge peak (see Ref.\ \cite{Willems17}). 

Figure \ref{sqfig} shows the experimental data on $S_{\rm{norm}}(F)$ for X-ray irradiated multilayer system, retrieved from Ref.~\cite{Andre21} (blue points), with an exponential fitting function, $S_{\rm{norm}}(F) = F \cdot \exp(d + c \cdot F)$ (orange dashed line), yielding the coefficients, $c=-0.035$ and $d=-0.034$ (cf. \cite{Andre21}). The experimental data are compared to the theoretical prediction for $S_{\rm{norm}}(F)$ obtained with the XSPIN model (black solid line). The prediction takes the interlayer electron transport into account. Note that the calculation of $S_{\rm{norm}}(F)$ for the theoretical predictions involves the multiplication of the theoretical raw signal $S(F)$ by the factor, $F_0/S(F_0)$, similarly as it was done for the experimental data in \cite{Andre21}. Therein, $F_0 \approx 0.4$ mJ/cm$^2$. The error bars plotted weight the theoretical results with the relative experimental error, calculated for the factor, $F_0/S(F_0)$, from the experimental data. The calculation used specifically the experimental errorbars for the fluence, $F_0$, and for the normalized scattering efficiency, $I_{\rm{eff}}$ (\cite{Andre21}; Fig. 2 therein). The error bars give an estimate for the uncertainty of the applied signal scaling. Note that the experimental data and theory predictions lay within the errorbars. 

In the plot, we also show the $S_{\rm{norm}}(F)$ obtained, when assuming a constant magnetization of the sample at all times, i.e., $M(t)=M(0)$. It starts to deviate from the normalized signal including demagnetization already for  fluences of a few mJ/cm$^2$, indicating the onset of the demagnetization contribution. This observation is in agreement with the trend seen in the experimental data. 

Generally, the limited range of fluences available for the actual experimental points (up to $\sim 40$ mJ/cm$^2$) and the large errorbars do not allow to accurately extrapolate the data towards higher fluences. For this purpose, an additional measurement of magnetic signal at higher fluence values would be helpful. However, one can expect that at very higher fluences, when structural damage strongly affects the sample, the magnetic scattering signal should ultimately disappear. Such behaviour has been experimentally observed at high X-ray fluences in \cite{Mueller13}. 

In any case, it should be emphasized that the region of interest for potential practical applications of X-ray induced demagnetization lies below the structural damage threshold. In this region, the  demagnetization is a reversible process, i.e., the magnetization of the sample, after a certain recovery time needed for excited electrons to relax, recovers its equilibrium value. The material can then be demagnetized with X rays again. With this analysis, we have shown that electronic processes strongly influence magnetic properties of the sample in this regime. In particular, our XSPIN model that treats electronic damage processes was able to describe the actual experimental data \cite{Andre21} with a good accuracy. The model can then be applied for predicting responses of various magnetic samples to X-ray pulses. Such study would be a significant step towards understanding and controlling X-ray induced femtosecond demagnetization in magnetic materials. 

%%%%%%%%%%%%%%%%%%%%%%%%%%%%%%%%%%%%%%%%%%%%%%%%%
\section*{Discussion}

With the theoretical model, XSPIN, we followed magnetic properties of X-ray irradiated magnetic multilayer system. We have shown that the demagnetization of such system, induced by X-ray pulses of fluences below the structural damage threshold, follows as a result of electronic damage processes. During the electronic relaxation, the occupations of magnetically sensitive levels in ferromagnetic Co transiently change, resulting in the  ultrafast decrease of Co magnetization. Within tens of femtoseconds, the magnetization reaches an equilibrium value, which remains stable on hundred femtosecond timescale. When one increases the fluence of X-ray pulse, the magnetization decreases to a lower  final  value.  This is reflected by the decrease of the time-integrated magnetic scattering signal with  increasing X-ray fluence. However, the timescale of the magnetization decrease is not affected by a change of pulse fluence. 

Further, we have shown that a similar behaviour of magnetic scattering efficiency can be observed for experimentally investigated multilayer systems. The electronic damage within the system layers is additionally influenced by pulse intensity attenuation and interlayer electron transport, which our model takes into account. Good agreement of our predictions with the data from the experiment by Kobs et al. \cite{Andre21} (within the limits of experimental accuracy) confirms the fidelity of this physical picture.  However, for ultimate model validation, it would be necessary to compare the XSPIN predictions on transient XUV induced magnetization to the respective time-resolved XUV experimental data, such as those obtained in \cite{Beaurepaire,Pfau12} for NIR induced demagnetization. An experiment performed for a single Co layer: (i) with magnetization aligned by an external magnetic field, (ii) then pumped with XUV rays, and (iii) probed with NIR pulses at varying time delays, would enable to collect such time-resolved data on magnetization which could then be compared to the XSPIN predictions. Such comparison would also require a significant extension of the XSPIN code, in order to take into account possible effects of long-range electron transport, as indicated in \cite{Pfau12}. This would be challenging, in particular, due to large spatial scales involved in the transport modeling. They would make  the simulation computationally expensive.

Up to our knowledge, the XSPIN model is the first model which couples a comprehensive quantitative description of X-ray induced electronic damage processes in solids, checked by earlier comparisons of its predictions with several experimental results on non-magnetic systems, with the description of transient magnetic processes  in solids. An earlier  theory model \cite{Scherz} ascribed the decrease of the magnetic scattering signal (tuned to the Co L$_{3}$ resonance) to the existence of a stimulated elastic scattering into the forward direction \cite{Wu}. This  mechanism has not been confirmed by later experiments  tuned to the Co M resonance \cite{Andre21,Schneider2020}.

Let us emphasize that the current model does not claim an immediate applicability for magnetic samples irradiated with {\it infrared radiation}. Different processes, acting on different timescales, can lead to demagnetization. In the X-ray/XUV regime and at 100 fs timescales, the electronic damage seems the fastest process which can drive it. It changes the magnetic state of the sample on a femtosecond timescale before an onset of any other  -- slower -- process which could otherwise demagnetize the material.  

The predominant role of electronic damage for X-ray induced demagnetization, confirmed with our theoretical study, opens a promising prospect for ultrafast demagnetization control. We can now predict with XSPIN how to affect the magnitude and timescale of the demagnetization by adjusting X-ray pulse parameters (wavelength, pulse duration and fluence), as well as by the choice of a magnetic material. Further experimental studies, supported by theory predictions with XSPIN, can then be performed, investigating the possible options for the control of demagnetization. This is a step towards prospective technological applications, e.g., XUV/X-ray light-controlled nanoscopic magnetic switches, operating on femtosecond timescales.

%%%%%%%%%%%%%%%%%%%%%%%%%%%%%

\section*{Methods}

%%%%%%%%%%%%%%%%%%%%%%%%%%%%%
\subsection*{Modeling interaction of X rays with solid  materials, using the code XTANT}

Modeling radiation damage in nanoscopic samples and solid materials has been performed for several years with various simulation techniques, e.g., \cite{XMDYN,ARGONNE,CRETIN, Medvedev_njp}. One of the tools is the hybrid code XTANT (X-ray-induced Thermal And Nonthermal Transitions) \cite{Medvedev_njp,Medvedev2017,Medvedev18,4open}. Using periodic boundary conditions, the  XTANT can simulate evolution of X-ray irradiated bulk materials. The code consists of a few modules dedicated to simulate various processes induced by the incoming X-ray FEL radiation: 

(a) The core of the XTANT model is a band structure module (in
\cite{Medvedev_njp,Medvedev2017,Medvedev18,4open} based on transferable tight binding  Hamiltonian, in \cite{dftbp22} replaced by the DFTB+ code \cite{dftbp20}), which calculates the transient electronic band structure of thermalized LEF electrons and the atomic potential energy surface. The latter also evolves in time, depending on the positions of atoms in the simulation box, and is used to calculate the actual forces acting on nuclei.

(b) After the forces act on atoms, the atoms move. Their actual positions are propagated in time, using a classical molecular dynamics (MD) scheme. It solves Newton equations for nuclei, with the potential energy surface evaluated from the band structure module. 

(c) Electron occupation numbers, distributed on the transient energy levels, are assumed to follow  Fermi-Dirac distribution with a transient temperature and chemical potential  evolving in time. The electron temperature changes  due to the interaction of band electrons with X-rays and  high-energy electrons; or due to their non-adiabatic interaction with nuclei (through electron--ion scattering \cite{Medvedev2017}). 

(d) Non-equilibrium fraction of electrons at high energies (HEF) and Auger decays of core holes are treated with a classical event-by-event Monte Carlo (MC) simulation. It stochastically models X-ray induced photoelectron emission from deep shells or from the valence band, the Auger decays, and the scattering of high-energy electrons.  In the code, at each time step an intrinsic averaging over $30000$ different Monte Carlo realizations of electron (and core hole) trajectories is performed, in order to calculate the average electronic distribution which is then applied at the next time step. 

(e) Electron--ion energy exchange is calculated, using a non-adiabatic approach \cite{Medvedev2017}. This energy is transferred to atoms by the respective velocity scaling at each MD step.

XTANT's hybrid approach enables computationally inexpensive simulations of relatively large supercells (containing up to 1000 atoms). The code treats all predominant excitation and relaxation processes within an X-ray FEL irradiated sample, including its non-equilibrium evolution stage, thermal and non-thermal processes, and structural transformations \cite{nikita2013prb}. In particular, all X-ray induced processes {\it exciting electrons} are taken into account in the model. Ballistic electrons are considered as high energy electrons. In the bulk material, they propagate with the restriction of periodic boundaries.
%%%%%%%%%%%%%%%%%%%%%%%%%%%%%%%%%%%%%%%%%%%%%%%%
\section*{Data Availability}
The data that support the findings of this study are available from the corresponding authors upon reasonable request.
%%%%%%%%%%%%%%%%%%%%%%%%%%%%%%%%%%%%%%%%%%%%%%%%%
%%%%%%%%%%%%%%%%%%%%%%%%%%%%%%%%%%%%%%%%%%%%%%%%
\section*{Code Availability}
The XSPIN code that supports the conclusions within this paper and other findings of this study is available under a license agreement. The licensor is Deutsches Elektronen-Synchrotron DESY, Notkestr. 85, 22607 Hamburg, Germany. Please contact the corresponding authors for more details.
%%%%%%%%%%%%%%%%%%%%%%%%%%%%%%%%%%%%%%%%%%%%%%%%
%%%%%%%%%%%%%%%
\section*{Acknowledgments}

We thank Christian Gutt, Dmitriy Ksenzov, Vladimir Lipp, and Jan Luening for helpful discussions. V. T., A. L., S. M., B. Z. acknowledge the funding received from the Collaboration Grant of the European XFEL and the Institute of Nuclear Physics, Polish Academy of Sciences. K. J. K. thanks the Polish National Agency for Academic Exchange for funding in the frame of the Bekker  programme (PPN/BEK/2020/1/00184). K. J. K. acknowledges also the CFEL-DESY Theory group for the hospitality during his six-month research stay in Hamburg in 2019-2020 financed by the National Science Centre (Poland) under the program SONATINA 1 no. 2017/24/C/ST3/00276. L. M. and A. P.-K. acknowledge funding by the Deutsche Forschungsgemeinschaft (DFG, German Research Foundation) -  SFB-925 - project 170620586.
%%%%%%%%%%%%%%%%%%%%%%%%%%%%%%%%%%%%%%%%%%%%%%%%
%%%%%%%%%%%%%%%%%%%%%%%%%%%%%%%%%%%%%%%%%%%%
%
\section*{Author contributions} 
%
%%%%%%%%%%%%%%%%%%%%%%%%%%%%%%%%%%%%%%%%%%%%

B. Z., A. L., and S. M. initiated this project; Further studies were lead with contributions of all 
authors; K. J. K. and V. T. performed the XTANT code extension to XSPIN -- in interactive discussion 
with B. Z., A. P.-K., L. M., F. C. -- and performed all calculations. All authors critically discussed the results and contributed to the manuscript, which initial version was written by K. J. K. and V. T.;  
K. J. K. and V. T.  contributed equally to this work.

%%%%%%%%%%%%%%%%%%%%%%%%%%%%%%%%%%%%%%%%%%%%
%
\section*{Competing interests} 
%
%%%%%%%%%%%%%%%%%%%%%%%%%%%%%%%%%%%%%%%%%%%%

The authors declare no competing interests.

%%%%%%%%%%
%\bibliographystyle{unsrt}
%\bibliographystyle{naturemag}
%\bibliography{references}

%\clearpage

%%%%%%%%%%%%%%%%%%%%%%%%%%%%%%%%%%%%%%%%%%%%%%%%%
\section*{Figure Captions}
%%%%%%%%%%%%%%%%%%%%%%%%%%%%%%%%%%%%%%%%%%%%%%%%%
%           FIGURES
%%%%%%%%%%%%%%%%%%%%%%%%%%%%%%%%%%%%%%%%%%%%%%%%%
%%%%%%%%%%% 
% FIGURE 1
%%%%%%%%%%
%%%%%%%%
% FIGURE 2
%%%%%%%%
%%%%%%%%
% FIGURE 3
%%%%%%%%
%%%%%%%%%%%%
%% FIGURE 4
%%%%%%%%%%%%
%%%%%%%%%
% FIGURE 5
%%%%%%%%% 
%%%%%%%%%%%%%%
%% FIGURE 6
%%%%%%%%%%%%
%% FIGURE 7
%%%%%%%
%%%%%%%%%%%%%%%%%%%%%%%%%%%%%%%%%%%%%%%%%%%%%%%%%
\end{document}